# Hairpins Participating in Folding of Human Telomeric Sequence Quadruplexes Studied by Standard and T-REMD Simulations

Petr Stadlbauer[1], Petra Kührová[2], Pavel Banáš[2], Jaroslav Koča[3,4], Giovanni Bussi[5], Lukáš Trantírek[3], Michal Otyepka[2], Jiří Šponer[1,3,*]

[1] Institute of Biophysics, Academy of Sciences of the Czech Republic, Královopolská 135, 612 65 Brno, Czech Republic
[2] Regional Centre of Advanced Technologies and Materials, Department of Physical Chemistry, Faculty of Science, Palacký University, tř. 17 listopadu 12, 771 46 Olomouc, Czech Republic
[3] CEITEC – Central European Institute of Technology, Masaryk University, Campus Bohunice, Kamenice 5, 625 00 Brno, Czech Republic
[4] National Center for Biomolecular Research, Faculty of Science, Masaryk University, Campus Bohunice, Kamenice 5, 625 00 Brno, Czech Republic
[5] Scuola Internazionale Superiore di Studi Avanzati, Via Bonomea 265, 34136 Trieste, Italy

* To whom correspondence should be addressed. Tel.: +420 541 517 133; Fax: +420 541 211 293; Email: sponer@ncbr.muni.cz

## ABSTRACT

DNA G-hairpins are potential key structures participating in folding of human telomeric guanine quadruplexes (GQ). We examined their properties by standard MD simulations starting from the folded state and long T-REMD starting from the unfolded state, accumulating ~130 μs of atomistic simulations. Antiparallel G-hairpins should spontaneously form in all stages of the folding to support lateral and diagonal loops, with sub-μs scale rearrangements between them. We found no clear predisposition for direct folding into specific GQ topologies with specific *syn/anti* patterns. Our key prediction stemming from the T-REMD is that an ideal unfolded ensemble of the full GQ sequence populates all 4096 *syn/anti* combinations of its four G-stretches. The simulations can propose idealized folding pathways but we explain that such few-state pathways may be misleading. In the context of the available experimental data, the simulations strongly suggest that the GQ folding could be best understood by the kinetic partitioning mechanism with a set of deep competing minima on the folding landscape, with only a small fraction of molecules directly folding to the native fold. The landscape should further include nonspecific collapse processes where the molecules move via diffusion and consecutive random rare transitions, which could, e.g., structure the propeller loops.

## INTRODUCTION

Telomeres are terminal regions of linear eukaryotic chromosomes. Their main function is to maintain genomic stability and facilitate chromosome replication. Telomeric DNA in vertebrates comprises of double stranded DNA consisting of tandem repeat sequence d(GGGTTA).d(CCCTAA), terminating in a single-stranded 150- to 250 nt long 3'-terminal G-rich overhang (1,2). Telomeres are shortened during every DNA replication (3,4) and their loss may eventually trigger apoptosis (5). A specialized enzyme, telomerase, helps to maintain telomere length (6). However, excessive telomerase activity may lead to cell immortality (7); cells in more than 80 % of cancers have indeed been shown to overexpress telomerase (8). The telomeric G-rich single-stranded overhang is capable of forming non-canonical evolutionary conserved structures called G-quadruplexes (GQ) (9), recently visualized *in vivo* (10,11), which may downregulate telomerase activity (12). Therefore, stabilization of telomeric GQs by small ligands may help in cancer treatment (8,13). GQ-forming sequences are widespread also in other parts of the genome and there is increasing evidence that formation of GQ structures may be a powerful way to regulate gene expression (14-19).

GQs are formed by stacking of guanine quartets. The quartet is a planar assembly of four *cis* Watson-Crick/Hoogsteen-paired (20) (shortly Hoogsteen-paired or *c*WH) guanines. Four carbonyl oxygens of each quartet aim towards GQ centre and form a central channel with a negative electrostatic potential. Monovalent cations are required to reside inside the channel to counterbalance the negative potential, and thus stabilize the GQ architecture. Mutual orientation of G-strands in quadruplexes is either parallel or antiparallel, and connecting loops can be lateral (edgewise), diagonal, or propeller (double-chain reversal), giving rise to parallel, antiparallel and hybrid GQ topologies. Antiparallel orientation of two strands requires that any two *c*WH base-paired guanines have different χ angle orientation, one being in *syn* conformation and the other in *anti*. Parallel strands impose the same χ angle orientation for base-paired guanines. This leads to interdependence between the *syn/anti* patterns in the GQ stems and the overall GQ topologies (21,22). Human telomeric sequence is well known for its structural polymorphism. Free energy computations suggest that polymorphism is a feature inherent to DNA GQs having odd numbers of guanines in the G-tracts, unless their topology is specifically dictated by their loops (23,24). The dominant GQ topology in a given experiment depends on exact nucleotide sequence including the flanking nucleotides, nature of stabilizing cation, DNA strand concentration, presence of cosolvents and perhaps on other factors (25-37). Six different topologies of human telomeric GQ have been elucidated by X-ray and NMR studies (30-36,38,39).

There are currently intense experimental efforts to understand the elusive process of GQ folding. Although earlier studies suggested that at least some GQ sequences may fold on a time scale of dozens to hundreds of milliseconds (40,41), most recent ensemble experiments indicate that the process of folding takes hours or even days (42,43). This is consistent with the latest single molecule experiments which discovered conformational states with similar lifetimes (44,45). The basic limitation of experimental investigations is that most of them do not allow confident determination of the structures that are populated in the course of the folding process, and thus plausible structural suggestions are sometimes based on intuition. The order parameters that are currently used to describe the folding processes may be not sufficient for the full characterization of folding. Some of them, e.g. derived from CD spectra, may occasionally be ambiguous (46). For example, although CD is commonly used to detect major changes of the GQ folded ensemble dominated by the basket-type GQ upon the $Na^+ \rightarrow K^+$ replacement (47), Raman spectroscopy (48) and NMR (49) revealed that sometimes the most populated fold may remain unaltered despite a profound change of the CD signal. Different and sometimes even kinetically unrelated ensembles may in principle overlap in the measured signals (50). Recently, first time-resolved NMR study of GQ folding intermediates for the human telomeric sequence emerged. The time resolution allowed to identify two competing long-lived structures corresponding to the hybrid-2 (off-pathway intermediate) and hybrid-1 (thermodynamic minimum) folds (43).

In fact, when considering all the currently available experimental data and their interpretation, the picture of GQ folding is far from being unambiguous. This may partially reflect limitations of the specific techniques. However, consistent interpretation of the experimental data may also be hampered by the fact that different experiments capture the folding process under different physical conditions, resulting, e.g., in genuine differences between ensemble and single molecule experiments. The unfolded (denatured) state ensemble, which significantly influences the folding process (45,51), depends on the specific physical process of unfolding. Thermally denatured, chemically denatured and low-entropy force-denatured ensembles are indeed not identical (51,52) and may give rise to diverse folding pathways.

Because of the genuine limitations of the experimental techniques, there have been increasing efforts to complement the experiments by modern computational techniques (53-62). The computational methods can provide in-depth insights into those parts of the folding process which are not within the resolution limits of the experiments (43). Among the available techniques, molecular dynamics (MD) simulation approaches are very promising, since they allow an optimal compromise between accuracy and computational efficiency. Standard MD studies attempting to mimic real thermal dynamics of DNA molecules are however limited by their μs timescale (63). This allows studying the properties of different structures (i.e., their associated local free energy basins) that can participate in the folding, but does not allow direct characterization of the full folding process. MD studies were performed for many GQ native structures (54,64-67), potential four-stranded intermediates with strand slippage (58,60) and G-triplexes (42,57,61). Simulation methods that enhance sampling can achieve larger conformational changes and even to certain extent sample the space between folded and denatured species. However, any acceleration of barrier crossing is associated with some drawbacks and may distort the picture of the folding process (63). The induced folding or unfolding might follow a pathway that depends on the exact protocol adopted to enhance sampling, which may complicate comparison with experiments made using specific control parameters and setups that are not equivalent to those in the computations (63,68). Nevertheless, even with enhanced sampling methods, it is unlikely we can achieve a complete folding analysis for systems that fold on such a long time scale as the GQs. Besides that, all theoretical studies are certainly affected by the approximations inherent to the simple molecular mechanical (force field) model (63,69).

The DNA GQ folding has been sometimes likened to funnel-like folding of small fast-folding proteins or RNAs, perhaps in light of some earlier experiments indicating fast folding. However, we suggest that there is a substantial difference between the principles of small protein and GQ folding. Short proteins have folding times in hundreds of μs to ms (51,70,71). Their sequences have been optimized to eliminate kinetic traps from their free energy surfaces, reducing their ruggedness to be around kT. On the contrary, the fact that the very short GQ-forming DNA sequences fold on very long time-scales indicates that GQ conformational surface is extremely rugged with deep competing basins of attractions (i.e., the substates) impeding fast spontaneous folding into the native basin of attraction (43,50). GQs have very limited sequence freedom for tuning the folding speed. Further, folding of short proteins is directed by formation of local secondary structure native contacts (71). This contrasts GQ DNA, where the most critical part of the folding may involve formation of the native consecutive ion-stabilized quartets. The quartets are based on non-local interactions and, in a given fold, must have a specific combination of *syn* and *anti* nucleotides (21,22). A decisive role of local contacts (which would include formation of the loop topologies) for the GQ folding is also not consistent with the fact that many GQs fold to diverse topologies upon subtle changes of the environment or flanking sequences. This indicates presence of several or even many competing basins of attractions. Then only a small fraction of the molecules reaches directly the thermodynamically most stable ensemble. The remaining molecules would be first trapped in competing substates, i.e. misfolded, resulting in extremely multi-pathway folding process. The native and most significant competing basins of attraction may be interchanged upon changing the experimental conditions, which may explain why the folding topology dominant in the final thermodynamics equilibrium is so sensitive to subtle changes of the folding conditions. Most molecules likely initially locate different (competing) folds and may need to unfold to make another folding attempt. This is consistent with our earlier simulation studies identifying numerous stable potential intermediates. It has been visualized experimentally for the folding of the human telomeric hybrid-1 type GQ for which the hybrid-2 structure has been shown as a major competing substate (43). The authors conclude that transitions between the two folds appear to be realized via the unfolded ensemble and note that the folding process can be affected by other substates which are below the resolution of the experiment. The number of substates that can be revealed by a given technique depends on its structural and time resolution.

Earlier, we studied potential four- and three-stranded architectures that may participate in GQ folding pathways (60,61). In this paper we complete the investigations by an extended set of simulations of potential double-stranded G-hairpin structures. G-hairpins are relevant not only to the earliest stages of the folding pathways (the hairpins were suggested to fold on the µs to ms time scale) (42,72), but also to various later phases of folding as the G-hairpins may be important parts of the ensembles during inter-conversions among different triplex and GQ arrangements (40,41,43,56,72-74). Only limited experimental and computational effort has been paid to G-hairpins so far (56,72-74).

We report three sets of investigations. First, a set of standard simulations is initiated assuming structures of the hairpins with *c*WH GG pairing with *syn-anti* orientations of the guanines corresponding to either different known GQ topologies or suggested GQ folding pathways. The aim is to characterize the conformational space associated with the native-like basins of the G-hairpins. The second set of standard simulations is initiated from unfolded single strand, in an attempt to initiate formation of a G-hairpin from a straight single strand. Finally, we use an extended temperature replica-exchange MD (T-REMD) (75) enhanced-sampling simulation to obtain a full-scale analysis of the conformational space of the G-hairpins that is independent of the starting structures.

**MATERIALS AND METHODS**

**Antiparallel hairpins with lateral (edgewise) loops**

We have examined three DNA sequences, differing in their 3'-terminal base: d[AGGGTTAGGG], d[AGGGTTAGGGT], and d[AGGGTTAGGGA]. The choice was based on the sequences of three known GQ NMR structures, basket-type 143D d[A(GGGTTA)$_3$GGG] (31), 3+1 hybrid-1-type 2GKU d[TT(GGGTTA)$_3$GGGA] (38), and 3+1 hybrid-2-type 2JPZ d[TTA(GGGTTA)$_3$GGGTT] (32). For the sake of simplicity, we do not list the flanking bases while the TTA loop is marked as '---' in the G-hairpin abbreviations used in the following text.

In antiparallel 5'-**G**GG---GG**G**-3' hairpins, the first G of the first G-stretch pairs with the last G of the second G-stretch (both in bold), etc. Since the guanines are *c*WH paired, the antiparallel arrangement implies that if one G in a base pair adopts *anti* orientation of its χ angle, the other G in the pair must adopt *syn* orientation (or *vice versa*) (21,22). Thus, if we assign specific order of *syn* (*s*) and *anti* (*a*) residues (henceforth '*s/a* pattern') to one G-stretch, the other adopts the inverted *s/a* pattern. Since there are three Gs per every G-stretch, there are $2^3$ = 8 possible *s/a* patterns for an antiparallel hairpin. We studied four of them: (i) 5'-*asa*---*sas*-3', (ii) 5'-*saa*---*ssa*-3', (iii) 5'-*ssa*---*saa*-3' and (iv) 5'-*sas*---*asa*-3' (Figure 1a). These four combinations maximize the number of *syn-anti* and *anti-anti* combinations of two consecutive Gs (the dinucleotide steps, 5'-GpG-3'), which are energetically favourable when participating in complete GQ stems (23,24). The other four possible *s/a* patterns have not been simulated due to the following reasons. The 5'-*aaa*---*sss*-3' and 5'-*sss*---*aaa*-3' patterns possess two unfavourable *syn-syn* steps (23,24). In addition, due to lack of any alternation of *syn* and *anti* in one G-stretch, such structures would be prone to vertical strand slippage (60,61). The 5'-*ass*---*aas*-3' and 5'-*aas*---*ass*-3' contain three unfavourable *syn-syn* and *anti-syn* steps (23,24).

Depending on the overall GQ topology, an antiparallel lateral loop hairpin can adopt either a narrow groove or a wide groove shape (Supporting Figure S1a-h) (21,22). We prepared both variants of the hairpins for all four studied *s/a* patterns, which means 4 * 2 = 8 hairpin variants (Figure 1a). When in addition considering the variants of the 3'-terminal bases (see above), we had altogether 24 starting structures. Eight of them could be directly prepared from the experimental structures (Figure S1a-h), using the first frames of the PDB files. The remaining structures were prepared manually by creating the missing 3'-terminal nucleotide and/or adjusting the *s/a* pattern. The built-up structures were subjected to extended equilibration to avoid possible instability introduced by the modelling procedure (see The Equilibration Protocol in the Supporting Information).

**Parallel hairpins with propeller (double-chain reversal) loops**

Propeller loop occurs when the two strands are parallel, which implicates that each base pair contains nucleotides with the same χ angle (21,22). The first G of the first G-stretch pairs with the first G of the second G-stretch, etc. We have taken three starting structures: (i) the first loop from 2GKU with the adjacent G-stretches with one T at both termini (d[TGGGTTAGGGT] with 5'-*saa*---*saa*-3' pattern; Figures 1b and S1i) (38), (ii) the last loop from 2JPZ with the adjacent G-stretches with one terminal residue at both termini (d[AGGGTTAGGGT]; 5'-*saa*---*saa*-3'; Figures 1b and S1j) (32), and (iii) the first loop from the parallel stranded X-ray structure 1KF1 with the adjacent G-stretches with one terminal residue at both termini (d[AGGGTTAGGGT]; 5'-*aaa*---*aaa*-3'; Figures 1b and S1k) (34).
    We have further examined parallel hairpin with single nucleotide propeller loop, namely, the last loop from the 3+1 hybrid structure of the Bcl-2 promoter GQ 2F8U with the adjacent G-stretches and one 5'-terminal T (d[TGGGCGGG]; 5'-*saa*-*saa*-3'; Figures 1b and S1l) (76). All propeller loops are associated with medium grooves (Figure S1i-l) (21,22).
    Note that the term 'parallel hairpin' is in the literature used also for a hairpin with topology resembling the antiparallel hairpin with a lateral loop, but with 5'-5' or 3'-3' chemical bond in the loop region. When referring to parallel hairpins in this article, we strictly mean a structure with parallel G-stretches linked by a propeller loop.

**Duplexes without loops**

To separate the effects of the loops, we extracted antiparallel wide and narrow groove duplexes and parallel medium groove duplex d[GGG]$_2$ from the 2GKU (38) structure by deleting the loops and all flanking residues. The antiparallel duplexes had the 5'-*saa*-3'/5'-*ssa*-3' pattern, while the parallel duplex had the 5'-*saa*-3'/5'-*saa*-3' pattern. The removal of loops introduces symmetry for antiparallel arrangements, so 5'-*saa*-3'/5'-*ssa*-3' is equivalent to 5'-*ssa*-3'/5'-*saa*-3'.

**Unfolded single stranded chains**

A single stranded d[GGGTTAGGG] B-DNA-like all-*anti* helix with the 5'-*aaa*---*aaa*-3' pattern was modelled using the NAB module of AMBER (77). Then, three other *s/a* patterns were prepared manually by flipping Gs into *syn* orientation, targeting the distribution of *syn* and *anti* Gs of consecutive G-stretches in the 2GKU (38) and 2JPZ (32) GQs, namely 5'-*saa*---*saa*-3' (to facilitate folding of the parallel hairpins with propeller loop), 5'-*ssa*---*saa*-3' and 5'-*saa*---*ssa*-3' (to facilitate folding of anti-parallel hairpins). We also built a d[AGGGTTAGGG] single-strand helix with 5'-*aaa*---*aaa*-3' pattern and from that we prepared a 5'-*asa*---s*as*-3' variant, corresponding to the lateral loops in the basket GQ 143D (31); the 5'-terminal A prevented *syn*-orientation of the first G (23,24).

**Triplex and quadruplex systems**

In order to test a specific hypothesis about folding of the basket structure which emerged from our interim results and our previous triplex study (61), we performed additional simulations of three hypothetical triplex intermediates **T1**, **T2** and **T3** with *c*WH paired Gs and of a hypothetical chair-type GQ marked as **Q1** (Figure 1c). We built up the d[A(GGGTTA)$_2$GGG] triplex **T1** with the 5'-*sas*---*asa*---*sas*-3' pattern and lateral (wide) → lateral (narrow) loop (groove) order, d[A(GGGTTA)$_2$GGGT] **T2** with 5'-*asa*---*sas*---*asa*-3' pattern and lateral (narrow) → lateral (wide) loop (groove) order, and d[A(GGGTTA)$_2$GGGT] **T3** with 5'-*asa*---*sas*---*asa*-3' pattern and lateral (narrow) → diagonal loop (groove) order. The GQ **Q1** had the d[A(GGGTTA)$_3$GGG] sequence, the 5'-*asa*---*sas*---*asa*---*sas*-3' pattern identical to the 143D GQ and lateral (narrow) → lateral (wide) → lateral (narrow) loop order. The **T1** fold was built based on the 2GKU structure by extracting the triplex with lateral (wide) → lateral (narrow) loop order and manual adjustment of the *s/a* pattern. The **T2**, **T3** and **Q1** folds were built from scratch, geometries of lateral loops were taken from the 2GKU. Cations were manually placed in their channels and all systems were subjected to prolonged equilibration (see Supporting Information). We performed three independent simulations of **T1** and **T2** and nine of **T3** (Supporting Information Table S1). The **Q1** GQ was simulated under standard conditions for 500 ns. The final structure was then utilized for two independent unfolding no-salt simulations. In no-salt simulations, no ions are present in the simulation box, including the GQ channel. Selected structures observed in one of the no-salt simulations were taken as starting structures for new standard simulations with ions (refolding simulations) (Table S1). Goal of this procedure was to observe possible structural rearrangements between **Q1** and the triplex structures. No-salt simulations represent an efficient tool to investigate likely rearrangements of GQ in early stages of unfolding and the subsequent refolding simulations aim to suggest movements that are relevant to late stages of folding, as in detail described elsewhere (60,78).

**Standard MD simulation**

All molecules were solvated by a truncated octahedral box of waters with minimal distance between the solute and the box border of 10 Å. We have used either TIP3P (79) or SPC/E (80) water models. These two water models differ somewhat in their kinetic properties (81), but this should not have any substantial systematic effect on our results, as confirmed by them. Thus, when a given system is simulated with both water models, all simulations are considered as equivalent. Solute molecules were first neutralized by counterions and then ~0.15 M excess salt was added, either NaCl or KCl. TIP3P-adapted and SPC/E-adapted Joung&Cheatham ion parameters were used (82). In some simulations, Amber-adapted Aqvist $Na^+$ (83) with Smith&Dang $Cl^-$ (84) ions in TIP3P water or Dang $K^+$ (85) with Smith&Dang $Cl^-$ ions in SPC/E water were also used, in order to eliminate a possibility that some detected ion binding sites are due to specific ion parameters (86). The no-salt simulations of the **Q1** molecule, one in TIP3P and one in SPC/E water box, were performed without presence of any salt. The residual charge was neutralized adding a homogenous compensating background.

Although the type of cation may affect final thermodynamics equilibrium of diverse folds in experiments, we can consider properties of $K^+$ and $Na^+$ ions as sufficiently similar in MD simulations, as discussed in ref. (78). Additional discussion about how to correctly compare simulations with experimental techniques can be found in ref. (87). The fact that $Na^+$ and $K^+$ can sometimes stabilize different folds under the condition of thermodynamics equilibrium does not mean that the other fold is entirely disrupted. Both folds are likely present with both cations, though the cations may modulate their relative free energies in such a way that one of them becomes undetectable. However, it does not significantly affect the kinetic stabilities and dynamics of both topologies on the µs time scale of the simulations (60). In other words, cation differences do not affect computations that do not directly investigate the free energy difference or transitions between the two topologies and with respect to the unfolded state. Thus, we do not expect any dramatic effect of the ion type or specific ion parameters on our results. When addressing specific properties where we suspected that the type of ion could affect our conclusions we made simulations with both $Na^+$ and $K^+$. Solvation, addition of ions and manual adjustment of *s/a* patterns were done in the xLEaP module of AMBER (77).

Parmbsc0 (88) + parm$\chi_{OL4}$ (89) + parm$\varepsilon\zeta_{OL1}$ (90) version (bsc0$\chi_{OL4}\varepsilon\zeta_{OL1}$) of the Cornell *et al.* force field (91,92) was used in almost all simulations (for parameters, see AmberTools15 May 2015 update or http://fch.upol.cz/en/ff_ol/index.php). The bsc0 modifies the *α/γ* balance to basically stabilize DNA simulations. The $\chi_{OL4}$ improves the balance of *syn/anti* nucleotide conformers and $\varepsilon\zeta_{OL1}$ modifies the backbone *ε* and *ζ* dihedrals. The $\chi_{OL4}$ and $\varepsilon\zeta_{OL1}$ adjustments improve simulations of GQs as well as B-DNA (89,90). In few simulations of the parallel stranded G-hairpins other force field combinations were also employed to test force field sensitivity of the results: bsc0$\chi_{OL4}$, bsc0$\varepsilon\zeta_{OL1}$, bsc0 alone or parm99 (93). The bsc0 alone was used for 6 standard simulations of unfolded single stranded chains, since they were done at the beginning of the project. We did not repeat these oldest simulations with the new bsc0$\chi_{OL4}\varepsilon\zeta_{OL1}$ version since their basic outcome was then confirmed by the extended T-REMD simulation using bsc0$\chi_{OL4}\varepsilon\zeta_{OL1}$. For overview of currently available nucleic acids force fields see ref. (63).

Electrostatic interactions were calculated using the particle mesh Ewald method (PME) (94,95), non-bonded cutoff was set to 9 Å. Temperature was held at 300 K and pressure at 1 atm using the Berendsen weak-coupling thermostat and barostat (96). The SHAKE algorithm was applied to all bonds involving hydrogens (97), and the integration step was set to 2 fs. The molecules were equilibrated as described in the Supporting Information. Production phase of the simulations was performed with GPU version of pmemd module of AMBER 12 (77). Resulting trajectories were processed in the ptraj module of AMBER and visualized in VMD .

**T-REMD simulations**

In order to enhance sampling of the d[GGGTTAGGG] single strand, we employed T-REMD (75). As starting structures, we utilized unfolded ssDNAs differing in orientations of the glycosidic torsions of all six guanines. Namely, we constructed all 64 possible *s/a* variants (considering six Gs: $2^6$ = 64) and these were taken as input structures for the T-REMD calculation involving 64 replicas. We checked that *syn* ↔ *anti* transitions were sufficiently sampled at all guanines and all replicas, so the final population of *s/a* variants was not biased by their distribution in the starting structures (Supporting Figure S2). Such choice of different starting structures should result in better convergence compared to the usage of one uniform starting structure for all replicas, though we in no case claim that our results are quantitatively converged. The TTA residues were in *anti* orientation in all starting structures.

The starting topology and coordinates of d[GGGTTAGGG] ssDNA were prepared using the tLEaP module of AMBER 12 (77). All structures were solvated using rectangular box with a 10 Å thick layer of SPC/E water molecules surrounding the solute (80). The T-REMD simulation was performed in ~0.15 M NaCl salt excess (84,98). All replicas shared the same number of water molecules, numbers of ions and size of the simulation box. All structures were minimized and equilibrated using equilibration protocol as described in Supporting Information. The T-REMD simulation was carried out using the AMBER suite of programs with bsc0$\chi_{OL4}\varepsilon\zeta_{OL1}$ (88-90). The temperatures of all 64 replicas spanned the range of 278-444.9 K, achieving exchange rates between 22% and 28%. The T-REMD simulation was performed at constant volume using the NVT ensemble in each replica, using PME with 1 Å grid spacing and 10 Å real-space cutoff. Since using the weak coupling approach (96) in combination with T-REMD can lead to artefacts (99), we used Langevin dynamics with friction coefficient of 2 $ps^{-1}$ as a thermostat in all replicas. Exchanges were attempted every 10 ps. Each replica was simulated to 1 µs, accumulating thus 64 µs in total. The trajectories were sorted either to follow continuous trajectories or to follow a given temperature in the ladder, and were analysed with the ptraj module of AMBER.

**RESULTS AND DISCUSSION**

**Antiparallel G-hairpins can readily participate in folding of GQs**

We have carried out series of standard 200-ns-long simulations of 24 different lateral-loop antiparallel G-hairpins using the bsc0$\chi_{OL4}\varepsilon\zeta_{OL1}$ DNA AMBER force field, in order to investigate their basic structural stability. We have carried out 6 independent simulations for each hairpin, resulting in 144 simulations with aggregate time 28.8 µs. Although we cannot claim that it provides a converged statistics, it should be sufficient to eliminate the most random results. Each simulation was carried out with different combination of water model/ion parameters. However, we consider all 6 simulations as equivalent, as we noticed no systematic effect of the solvent parameters on the simulation outcome. The simulation time scale was chosen to observe a representative spectrum of structural transitions and should be sufficient to decide if a given hairpin is intrinsically stable enough (having sufficient life-time) to directly participate in GQ folding pathways. This part of our work essentially monitors the unfolding rate of the folded G-hairpins. Note that the life-time of the G-hairpins in the simulations may be underestimated by the force field and thus we estimate rather the bottom border of their lifetimes (61). Nevertheless, our results should correctly reflect their relative structural stability order.

The simulations suggest that antiparallel G-hairpins are stable enough to participate in the folding pathways of GQs. At the same time, the simulations reveal exceptionally rich dynamics including entirely stable simulations, various local perturbations, transitions between narrow and wide groove G-hairpins, spontaneous formation of diagonal loop G-hairpins, strand slippage movements, complete unfolding events and even rare re-folding events. The dynamics is sensitive to the flanking nucleotides.

The basic results are summarized in Tables 1-3. We divide the simulations into three groups: i) Simulations with at least two initial *c*WH GG base pairs present at the simulation end. This means either completely stable structures or simulations with a transient reversible visit of geometries different from the starting one. This group of simulations is referred to as *stable simulations* further in the text. ii) Simulations with *perturbed* (compared to the start) structures at the end, nevertheless still keeping an overall G-hairpin-like structure; these include diagonal G-hairpins with *trans* Watson-Crick/Watson-Crick (*t*WW) (20) base pairs, hairpins that underwent narrow → wide groove transition (or vice versa), locally misfolded hairpins with strand slippage, etc. iii) The last group includes *disrupted* (i.e., unstable, unfolded) structures, usually unwound ss-helices or coils. The first group of structures would be competent to promote folding to the desired topology within the context of the full sequence while the second group would need only minor rearrangements to do so. The third group is unsuitable for any immediate folding attempt. Altogether, 73, 38, and 33 simulations were classified as stable, perturbed and unstable, respectively.

**d[AGGGTTAGGG] antiparallel G-hairpin**

Table 1 summarizes the simulations of the d[AGGGTTAGGG] hairpin. Its narrow groove variant displayed remarkable stability for the naturally occurring 5'-*asa*---*sas*-3', 5'-*saa*---*ssa*-3', and 5'-*ssa*---*saa*-3' patterns, known from existing GQ atomistic structures. 17 simulations were stable, 1 simulation perturbed and none was unstable. Nevertheless, the structures were not rigid and exhibited rich local dynamics. The 'non-native' 5'-*sas*---*asa*-3' pattern was less stable, with 2 simulations entirely disrupted and 1 simulation perturbed.

The wide groove d[AGGGTTAGGG] hairpin was visibly less stable for all four studied *s/a* patterns. The 5'-*saa*---*ssa*-3' pattern had a tendency to form *t*WW GG base pairs. Still, 2 of its 6 simulations resulted in a complete structure loss and 1 into a modest perturbation. The remaining three *s/a* patterns entirely unfolded in 3-5 of the 6 individual simulations (Table 1).

**d[AGGGTTAGGGT] antiparallel G-hairpin**

Among the narrow groove hairpins, only the 5'-*asa*---*sas*-3' pattern remained completely stable in all 6 simulations (Table 2). The other narrow hairpin *s/a* patterns showed tendency to change the *c*WH GG pairing into the *t*WW diagonal G-hairpin pairing, in 12 out of 18 simulations. In 3 simulations of the 'non-native' 5'-*sas*---*asa*-3' pattern, the rearrangements proceeded further into the wide groove G-hairpin (Figure 2), i.e., we evidenced a spontaneous rearrangement of the structure from narrow to wide G-hairpin through a diagonal hairpin. Important factor affecting the simulations was a formation of the terminal canonical *c*WW AT base pair between the first and last nucleotide of the whole sequence. This event always preceded the transition from *c*WH into *t*WW GG pairing; the stable 5'-*asa*---*sas*-3' narrow groove pattern never formed the terminal AT base pair. The role of the AT base pair is a nice example how a specific interaction with non-G bases can modulate structural dynamics of G-hairpins at the atomistic level, in this particular case the balance between narrow groove, diagonal and wide groove antiparallel hairpins.

For wide groove d[AGGGTTAGGGT] hairpins, strand slippage was observed in 9 out of 24 simulations. The structures slipped towards the 5'-end can be stabilized by formation of a base pair between the first G of the second G-stretch (G8) and the first T of the loop (T5) (Figure 3) or a base-phosphate interaction between amino group of G8 and the T6 phosphate group. Out of the four *s/a* patterns, the 5'-*saa*---*ssa*-3' variant was the most stable, with no unfolding and only 1 case of strand slippage. The patterns 5'-*asa*---*sas*-3' (the most stable one in the narrow variant, see above) and 5'-*sas*---*asa*-3' were the least stable (Table 2).

**d[AGGGTTAGGGA] antiparallel G-hairpin**

The narrow groove 5'-*asa*---*sas*-3' and 5'-*ssa*---*saa*-3' patterns were the most stable, each of them in 5 simulations (Table 3). They formed stacking of the terminal As on each other and on the terminal GG pair. The other two *s/a* patterns were perturbed or stable to approximately the same extent. Formation of the diagonal loop with *t*WW GG pairing was observed in 8 simulations, almost evenly distributed between the *s/a* patterns. One large-scale refolding event after initial structure loss was observed (see below). The wide groove hairpins exhibited roughly equal stability as in case of the d[AGGGTTAGGGT] hairpin. The 5'-*saa*---*ssa*-3' pattern never unfolded, with formation of *t*WW GG pairs in half of the simulations. The other three *s/a* patterns unfolded in 2 simulations each.

**Set of interactions stabilizing the 5'-*xxa*---*sxx*-3' lateral-loop patterns**

The simulations suggest that, regardless of the flanking residues, the non-native 5'-*sas*---*asa*-3' lateral loop pattern suffers from opening of the GG base pair adjacent to the loop (the closing pair) while the native patterns are stable. In case of the narrow hairpins, it can be explained as follows. If the first G after the loop is *syn*-oriented, it usually forms intra-nucleotide hydrogen bond between its amino group and its phosphate group (Figure 4). This can stabilize all 5'-*xxa*---*sxx*-3' patterns, but none of the 5'-*xxs*---*axx*-3' patterns. This hydrogen bond is usually present in full GQs with narrow grooves (e.g. in 143D, 2MBJ, 2HY9, 2GKU and 2JSM) and can be considered as a native GQ interaction. Such intra-nucleotide base-phosphate interaction sometimes appears, albeit with much lower population, also within the second or terminal *syn* G in the second G-stretch.

Further stabilization of the 5'-*xxa*---*sxx*-3' patterns may come from two additional hydrogen bonds of the first T, namely between T(O4) and the *anti*-oriented G(N2) preceding the loop, and between the T(O2) and the loop A(N6) (Supporting Figure S3). These two H-bonds are probably relatively weak because they are not fully linear. They are rather rarely seen in the experimental structures, e.g., in 143D and 2JSM. This may be due to the fact that the A is, in fully folded GQs, often paired with bases from other loops. Nevertheless, the shape of the lateral loop in our stable G-hairpin simulations is very similar to the shape of the loops associated with narrow grooves of almost all human telomeric GQs experimental structures, except of 2JPZ. Perhaps, the weak T(O4)-G(N2) and T(O2)-A(N6) hydrogen bonds seen in our G-hairpin simulations are disrupted upon formation of more stable native interactions in later GQ folding stages. The structure of the loop is also supported by stacking of the middle T with the A (Figure S3), which is seen also in complete GQs. Thus, some 'native' G-hairpins may be weakly favoured during the folding and some of their native interactions may locally influence the folding free energy landscape in the early stages of folding. Such specific molecular interactions may contribute to the GQ topology rules, for a recent review see ref. (22).

**Cation binding in antiparallel G-hairpins**

The simulations reveal notable cation binding sites around the G-hairpins. Two sites are associated with the -*sa*- GpG steps: a carbonyl binding site in both narrow and wide hairpins, equivalent to the channel binding site between two quartets in GQs, and a narrow groove binding site, where the cation is located between phosphates in the narrow groove (Figure 4). We have already observed the latter binding site in G-triplex intermediates (61) and in long simulations of GQs (unpublished data). Occupation of both sites by $Na^+$ is roughly equal when the -*sa*- GpG step is adjacent to the loop. The narrow groove binding site is weaker when the -*sa*- GpG step involves the terminal GG pair, perhaps due to greater fluctuations of the terminal nucleotides affecting spatial orientation of the phosphate groups. Potassium cations were attracted to the carbonyl binding site more than to the narrow groove site, with Dang parameters leading to the weakest groove binding. This behaviour could be attributed to smaller radii of the $Na^+$ compared with $K^+$. Nonetheless, the cation binding does not seem to be critical for stability of narrow hairpins. We observed stable narrow hairpins in the no-salt simulation of the hypothetical **Q1** chair GQ reported below.

No significant groove binding was observed for wide hairpins and diagonal hairpins with *t*WW base pairs, if formed in the simulations. All antiparallel hairpin geometries, narrow, wide, diagonal and even misfolded with slipped strands could temporarily bind cations in the loop region, depending on the actual spatial arrangement of bases in the loop. In summary, the simulations did not find any major difference between NaCl and KCl simulations which could explain the experimentally known effect of the ions on the equilibrium GQ structures. However, this result can be affected by the short time-scale of our simulation and inability of the force field to capture the difference. In addition, it is possible that the effect is not deducible from properties of isolated G-hairpins. Perhaps, the subtle differences between the $Na^+$ and $K^+$ when bound to the narrow groove and carbonyl binding sites may potentially contribute to the sensitivity of the thermodynamically preferred GQ topologies to the type of the ion. However, quantifying this issue would require different types of computations and would be exceptionally difficult.

**Why is the first G in a narrow groove following a lateral loop usually *syn*?**

As noted above, antiparallel hairpins with the naturally occurring *s/a* patterns were found to be more stable than the 'non-native' variant. This might indicate that they are better poised to serve as intermediates in GQ folding pathways (40,41,56,72,73). Possibly, their higher intrinsic stability may even be transferred to the full GQ structures. The observations of cation binding support previous calculations predicting the binding site on the carbonyl side of -*sa*- GpG steps to be preferred over the -*aa*-, -*ss*- and -*as*- steps (56). It is known that the first nucleotide after narrow lateral loop in GQs is usually *syn* (22). We speculate that it could be explained by the potential of the -*sa*- GpG steps to bind cations in narrow groove, the above-explained capability of *syn*-oriented Gs to form internal hydrogen bond between the amino and phosphate groups and perhaps formation of the two above-described weaker H-bonds within the loop.

**Diagonal antiparallel G-hairpins are viable intermediates**

Diagonal hairpins are antiparallel hairpins with *t*WW GG base pairs. Although all our simulation started with lateral G-hairpins with *c*WH GG base pairs, we observed spontaneous sampling of the diagonal G-hairpins in many simulations (Tables 1 – 3). Detailed population of different hairpins is given in the Supporting Information Appendix Tables SA1 – SA24. The diagonal G-hairpins compete mainly with the wide groove G-hairpins, though we have also seen simulations where diagonal hairpins served as intermediates in the interconversion between narrow and wide groove hairpins (Figure 2); both directions were sampled in the whole simulation set. Thus, there has been no need to start additional simulations using the diagonal *t*WW G-hairpin arrangement. We conclude that antiparallel G-hairpins can readily support formation of narrow lateral, wide lateral and diagonal loops, allowing easy rearrangements between them. All these G-hairpins can easily participate in GQ folding and rearrangements. Role of diagonal hairpins in folding of the hybrid GQ has been suggested (43).

**Parallel G-hairpins with propeller loops are entirely unstable in simulations**

Out of the 19 individual 200 ns simulations of parallel G-hairpins with TTA loops, 18 unfolded; 9 even within the first 10 ns. The behaviour was independent of the used force field version (Supporting Table S2). The most common unfolding pathway started with counter-rotation of the strands into structures with roughly perpendicular G-stretches (Figure 5). Then, it proceeded into various coil or ss-helix geometries.

Once significantly unfolded, the parallel hairpin never reformed. All 5 simulations of the studied single-nucleotide propeller loop G-hairpin also unfolded.

When combining the present results with earlier studies of other systems (54,60,64,78,100) we see the following simulation behaviour of parallel G-hairpins. They are entirely stable in simulations of full GQs stabilized by the channel ions, even in absence of the bulk ions (78). However, they are very unstable in all other situations, namely, in simulations of GQs lacking the channel ions (54,60), in G-triplexes even in excess salt conditions (61), and in isolation (the present work). This indicates that the propeller loop structures are held by the GQ core and do not contribute to the stability of the GQ structure. This simulation picture indirectly contrasts with the abundance of the propeller loops in known experimental structures of GQs. The single nucleotide propeller loop is actually supposed to be one of the most stable loops (101-105).

In the whole simulation set we observed one transformation of a parallel G-hairpin with TTA propeller loop and one with the single nucleotide propeller loop into antiparallel arrangements. These transitions were not accompanied by any *syn* ↔ *anti* flips and thus did not result in perfect G-hairpins. The 5'-*saa---saa*-3' hairpin formed a slipped structures with two *c*WH GG base pairs. Such antiparallel hairpin could straightforwardly serve as an intermediate in formation of GQs with two quartets. The 5'-*saa-saa*-3' hairpin adopted an antiparallel structure stabilized by *t*SS-like (*trans* sugar edge/sugar edge) GG pair closing the loop, *t*WH GG base pair in the middle and terminal *t*WW GG pair (Supporting Figure S4). These two simulations directly visualise that the parallel structures are less stable than the antiparallel arrangements.

The reasons of the instability of the parallel G-hairpins in simulations are unclear. One option is that it is a genuine property of these hairpins. However, then their common presence in the experimental structures would be non-trivial, and their formation could be one of the bottlenecks of the folding attempts of the individual molecules. Alternatively, some approximations of the force field could under-stabilize the propeller loops in simulations. It could be caused by inaccurate description of the backbone substates needed for the chain-reversal topology (100) or of the interactions between the closely spaced phosphate groups. Although the DNA backbone dihedral potentials have been extensively optimized in the past, the simple pair-additive DNA force field model remains imperfect, as demonstrated by comparison with quantum chemical benchmarks (24,106). Our ability to tune the force field by simple refinements of the essentially unphysical dihedral potentials should not be overrated (63).

**Absence of loops introduces instability**

Simulations of the *c*WH paired d[GGG]$_2$ duplexes without any loop and flanking residues are unstable, regardless of being parallel or antiparallel (Supporting Table S3). Perhaps, instability of the *c*WH d[GGG]$_2$ duplexes is promoted by introduction of another DNA terminus by the loop excision, which increases susceptibility of such a short duplex to unfold due to simulation end effects (107). Nevertheless, instability of d[GGG]$_2$ duplexes does not pose any problem to construct the intramolecular GQ folding pathways. Further details are described in Supporting Information. A plausible model of formation of intermolecular GQs lacking loops does not require *c*WH paired duplexes as intermediates (58).

**Single stranded helices tend to form misfolded antiparallel hairpins in standard simulations**

Four standard simulations of d[GGGTTAGGG] and two of d[AGGGTTAGGG] starting from the unfolded (straight) single stranded helix with different *s/a* patterns were done (Supporting Table S4). They exhibited a tendency to bend towards the antiparallel-hairpin-like arrangements. The simulations resulted into a variety of misfolded structures with strand slippage. None of them was a properly folded hairpin with three GG pairs and a TTA loop, even though some of the helices started with manually pre-ordered self-complementary *s/a* GQ pattern. Nevertheless, the simulation with 5'-*saa---ssa*-3' pattern attempted a move towards the right G-hairpin. It has formed the lateral TTA loop with the correct adjacent *c*WH GG base pair. However, further progression of the folding was blocked by stacking and base-phosphate interactions of the other Gs (Supporting Figure S5). This is an atomistic example how non-native interactions enhance ruggedness of the folding landscape. The initial pre-folding event occurred at 900 ns. We have prolonged the simulation to 1500 ns, but no further development has been seen. Further details about the simulations are given in Supporting Information.

**Rare refolding events observed in simulations of the antiparallel G-hairpins**

As explained above, a few dozens of simulations of antiparallel G-hairpins resulted into a complete unfolding of the initial structure. We observed two cases of spontaneous re-folding, when essentially the starting conformation has been re-established after entire unfolding. Such simulations provide unique atomistic insights into the potential G-hairpin folding pathways and are in detail described in Supporting

Information. However, we still caution that both re-folding events might have been affected by the structural memory effect and the molecules might have been bouncing back from the transition state ensemble without reaching a true unfolded state ensemble (108).

**T-REMD simulation of d[GGGTTAGGG] reveals enormous richness of its conformational space with a clear trend to form antiparallel hairpin-like structures**

Since standard simulations were not efficient to study folding of the G-hairpins, we attempted an enhanced sampling method. We have selected the temperature-accelerated T-REMD approach, a robust method often considered as benchmark for small systems, such as DNA hairpin loops with canonical B-DNA stem (109,110), UNCG and GNRA RNA hairpin loops (111,112) and RNA tetranucleotides (113,114). Although the T-REMD method does not allow derivation of kinetics of the folding, the reference replica at a given temperature should give the correct thermodynamics (relative population of different species), provided the simulation is converged.

Our T-REMD run simulated 64 independent replicas, each for 1 μs. It has been initiated from a straight ssDNA helix, with each replica initially adopting one of the 64 possible *s/a* patterns of the G-stretches, to sample distinct regions of the conformational space since the beginning of the simulation. We carefully monitored all 64 replicas at all temperatures, however, we decided to not report all populated structures in detail. It would mean to deal with hundreds of different structures, which is not necessary for the purpose of the present work.

Although the simulation was certainly still far from been rigorously converged, we have achieved broad sampling. The fully paired antiparallel G-hairpins were spontaneously reachable, but their population was only about 0.01 % (see below). It may reflect complexity of the conformational space (i.e., the T-REMD was still too short), some underestimation of the stability of the ideally folded G-hairpins by the force field (61) or real absence of such structures as dominantly populated species in the equilibrium ensemble. It will be interesting to see if future experiments directly visualize formation of well-paired G-hairpins or rather reveal a complex mixture of diverse structures similar to that predicted by the T-REMD run. Until today, the most direct evidence of formation of stable G-hairpins came from DNA origami experiments by Sugiyama's group (73) and NMR experiments by Plavec's group (115). The first method, however, did not have atomistic resolution while the NMR study predicted dimerization of diagonal hairpins for the *Oxytricha nova* sequence d[$G_4T_4G_4$]. The longer G-tracts and the dimerization may bring additional thermodynamics stabilisation to the G-hairpins.

Nevertheless, the most important result was that our T-REMD easily reached structures resembling antiparallel G-hairpins, except that they were not fully paired (see below). These structures actually formed ~17% of population of the reference 278 K replica. Therefore, a broad spectrum of structures that can support lateral and diagonal loops is easily accessible. On the other hand, we have seen absolutely no tendency to form any structures that could lead to parallel G-hairpins with propeller loops; not a single such fluctuation was detected. All these results were consistent with the preceding parts of the paper.

Formation of rather compact bent anti-parallel conformations was a hallmark of all replicas, with a slightly higher propensity of high-temperature replicas to adopt straight conformations. In order to monitor the global bending, we measured the end-to-end distance, i.e., the distance between centres of mass of the terminal guanines. Figure 6 summarizes the population of the bent and straight conformations in the reference 278 K replica as well as the populations summed over all temperatures. The minor population of end-to-end distance of 28-30 Å corresponding to the straight single strand conformation (region III in Figure 6) was well separated from the other geometries. The other structures were various types of bent conformations. They included conformations with stacked terminal Gs, either similar to the GQ-like hairpin helix with stacked rather than base-paired terminal guanines or a ring-like conformation with two bends and stacked terminal bases (region I in Figure 6), different types of base-pairing of the terminal guanines (region II in Figure 6) and conformations, where the terminal Gs did not directly interact (also region II in Figure 6) such as structures with inter-strand stacking in hairpin-like conformation with shifted base pairing within the stem. Further analysis (see Supporting Information) revealed high propensity of bending in the TTA region at low temperatures, while at high temperatures the preference was less pronounced (Figure 7). Formation of the *t*WW or *c*WH GG pair (both wide and narrow groove geometry) adjacent to the loop was found to be the most frequent out of all possible GG base pairs in the reference replica (Supporting Table S5), with 17.3% population. All such structures can be considered as the first step towards forming the antiparallel G-hairpins, although further zipping to two and three base pairs was seen with only ~0.16% and ~0.01% population, respectively.

Our T-REMD achieved sufficient amount of *syn ↔ anti* transitions (Supporting Figure S2), in contrast to the standard simulations. REMD accelerated the sampling by temperature, so at higher temperatures, it is more likely to overcome the potential energy (enthalpic) barriers and achieve *syn ↔ anti* transitions. The total population of *s:a* orientations was almost uniformly ~40:60 for G2, G3, G7, G8 and G9. The 5'-terminal guanine G1 revealed swapped preference favouring *syn* orientation in rate ~80:20, stabilized by a 5'-OH…N3 intramolecular hydrogen bond (23,24). Using these populations, we can calculate idealized populations of all 64 *s/a* patterns and compare them with the actual T-REMD results. Idealized populations would be achieved when the guanines sample the *syn* ↔ *anti* transitions independently. We indeed found that with increasing temperature of the replicas, the observed *s/a* patterns converged to the ideal populations. So, at high temperatures the *s/a* orientations of Gs were mutually independent and mostly driven by entropy, guaranteeing proper sampling of the pool of all *s/a* combinations. On the other hand, we found several *s/a* patterns that were significantly more populated at low temperatures compared to the idealized populations, e.g., 5'-*sss*---*asa*-3', 5'-*sas*---*saa*-3', 5'-*saa*---*saa*-3' and 5'-*sas*---*asa*-3' (Supporting Table S6). Interestingly, only the last *s/a* pattern could form an antiparallel hairpin compatible with a three-quartet GQ. We analysed clusters of the base-pairing interactions for these four *s/a* patterns and we indeed found that the anomalously well populated *s/a* patterns could be explained by specific interactions and structures (Supporting Figure S6). We describe further details of the T-REMD simulation in Supporting Information.

**Hypothetical idealized basket-type GQ folding pathway**

In the above paragraphs we emphasized that the simulations suggest that GQ folding is an extremely multi-pathway process over exceptionally rugged free energy surface. In this paragraph we show that the simulation results can be used to suggest idealized (simplified, few-state) pathways in a similar manner as when interpreting experiments. We construct a pathway from single strand to three-quartet basket-type topology lacking the problematic propeller loops. We simulated set of triplex structures and a hypothetical three-quartet chair-type GQ, marked as **Q1**. **Q1** has never been observed in atomistic experiments, but it may be populated during the folding. The model is summarized in the Figure 8. We have simulated **Q1, T1, T2** and **T3** structures. **T4** has been spontaneously formed. **143D** is the observed GQ and properties of structures **H1** – **H24** can be deduced from the simulations of the G-hairpins. **Q2** is a hypothetical basket variant with opposite strand progression compared to **143D** and we try to explain why it is not formed.

The simulations of **T1** displayed tendency to change the lateral wide loop into a diagonal loop. After two microseconds, two simulations revealed two triads rearranged in such a way that one loop shifted from the initial lateral wide into diagonal arrangement. The triad adjacent to the loop adopted an intermediate triangle geometry (61) in one simulation and did not change in the other simulation (Supporting Figure S7). This transition is consistent with our previous triplex study (61). **T2** remained stable in two simulations and in one simulation one G-stretch detached while the remaining narrow groove hairpin stayed stable. In contrast, **T3** exhibited instability of its diagonal loop in all nine simulations. In three of them, one G-stretch unwound so that the narrow groove hairpin remained. In the remaining six simulations the triplex partly or fully converted its diagonal loop into lateral wide loop, leading to the **T3** → **T2** conversion. As the A1-T11 base pair adjacent to the diagonal loop (see Methods for initial construction of the structures) in the starting structure could destabilize the diagonal loop, we have manually disrupted the base pair in six simulation. However, the **T3** diagonal loop remained unstable. We speculate that the **T3** triplex is less stable due to a less favourable topology of the backbone, as in **T3** the TTA loop starts from the narrow groove while in **T4** (and **143D**) it emerges from the wide groove; these two arrangements are topologically different. Although we cannot rule out some other reasons of **T3** instability, it seems that the **T3** triplex is less stable than the other triplexes and tends to convert to **T2**.

Since the hypothetical **Q1** chair GQ is stable in standard simulations with full ion binding, we applied the unfolding (denaturing) no-salt simulation (see refs. (60,78) for explanation of the methodology). In the first simulation the **Q1** was split into a combination of two narrow hairpins **H12**. The second simulation resulted into triplex-like intermediate. We then tried several standard excess salt simulations, in order to stabilize the perturbed structures. Four of them continued to unfold, one refolded back into **Q1**, and one converted into the **T4** structure, i.e., in a direction towards the **143D** basket.

When combining these results with the data for the antiparallel G-hairpins, we construct the following scenario for folding of the basket **143D** GQ (Figure 8) while avoiding the alternative **Q2** structure. The sequence of the GQ **143D** is d[A(GGGTTA)$_3$GGG] with the 5'-*asa*---*sas*---*asa*---*sas*-3' pattern. Considering stability of the narrow hairpins **H1** and **H2** in the simulations of d[AGGGTTAGGGT] and d[AGGGTTAGGG] with the 5'-*asa*---*sas*-3' pattern, we suggest that these two G-hairpins are stable enough to allow further folding with a sufficient probability. If both hairpins are formed simultaneously, they can merge together to form the chair-like GQ **Q1**. When only one of the two hairpins is present, it can bind an additional G-stretch with a compatible *s/a* pattern to form either **T1** or **T2** triplex. Contribution of wide hairpins of these two sequences and *s/a* patterns to folding is less likely due to their low stability, revealed in the first part of this study. Thus, we speculate that all intermediates containing the wide hairpin variant, including **H13** and **H24**, are less probable. Because the **T3** triplex is unstable and spontaneously changes into **T2**, neither **T3** nor **H24** can proceed to the GQ **Q2**. **Q2** actually contains the same unstable lateral narrow → diagonal loop order which makes **T3** unstable. On the other hand **T1** eventually changes its lateral wide loop into diagonal, giving rise to **T4**. **T4** can then pick up the last free G-stretch to form the stable **143D** GQ. **T1** and **T2** can interconvert via unfolding or via the **Q1** GQ, so it is likely that **T2** either gradually unfolds or forms **Q1** and then follows the track of **T1** (Figure 8). The model explains why **143D** is formed more likely than **Q2**.

**Comment on propeller loops**

Idealized folding pathways of the hybrid GQ topologies proposed in the literature usually form the propeller loop at the very end (41,42,54,56,116), which is in accord with their instability observed in simulations here and earlier (54,60,61). These pathways begin with formation of antiparallel hairpins and here we show that G-hairpins with *s/a* patterns corresponding to the hybrid GQs (wide groove 5'-*saa---ssa*-3' and narrow groove 5'-*ssa---saa*-3') are stable enough to participate in the folding process. More problematic could be formation of the all-parallel GQ with three propeller loops. Perhaps, it could be structured using a mechanism suggested for intermolecular GQs, i.e., initial coil-like ensemble of four strands, formation of first ion-stabilized quartet, and subsequent structuring of the molecule via strand slippages (58).

**Comment on the *Oxytricha nova* GQ**

Common models of GQ folding assume consecutive additions of G-strands. Likewise, most of our calculations investigated possibilities of such scenarios. However, recent experiments by Plavec's group suggested yet another possibility, which can further complicate the folding landscape. They have suggested a prefolded four-stranded structure in absence of the ions with two weakly interacting diagonal hairpins, which quickly converts into the GQ after adding ions. This has been proposed based on an NMR study for dimeric *Oxytricha nova* GQ d[$G_4T_4G_4$]$_2$ (115). Such intermediate also appeared in our preliminary simulation of no-salt induced unfolding of this GQ (data not shown). It would be exceptionally interesting to clarify if analogous prefolded structures may form also for systems having less than four quartets.

**Limitations of idealized folding pathways**

The folding model depicted in the Figure 8 (and similar pathways suggested in the literature) is simplified. Its entirely unrealistic prerequisite is that the *s/a* patterns of the approaching components are always the native ones. This is also assumed in the pathways presented in the literature (42,56). The **H1 → T1 → T4 → 143D** pathway without inclusion of any other stable intermediates is likely part of the real folding process, but ignores interference from off-folding intermediates with different *s/a* patterns, including alternative three- and two-quartet GQs. Complexity of folding emerges even for our simple model not allowing for any changes of the *s/a* pattern. The alternative **H2 → T2 → Q1 → T1** pathway contains the potential intermediate **Q1**, a chair GQ, which likely would be a long living intermediate slowing down the overall native GQ folding rate as a deep competing basin of attraction, i.e., a stable misfolded state.

As stated elsewhere, it is likely that instead of pure intermediates, such as **H1** or **T1**, complex ensembles of structures are present during folding (43,117,118). MD simulations show atomistic models of many stable misfolded hairpins, triplexes and GQs (60,61). It indicates that the real process is slowed down by numerous intermediate structures with variable populations and lifetimes. It is easy to imagine a misfolded GQ, which needs to unbind one G-stretch, flip *syn* and *anti* orientation of guanines and undergo another folding attempt, which might again lead to another misfolded GQ. Transitions between different GQ folds may require movements through the unfolded ensemble (43). Here we wish to point out that the differentiation between what is unfolded and what is misfolded structure may depend on the time and structural resolution of a given method. Let us assume binding of an unbound G-stretch to a triplex, in order to form a specific three-quartet GQ. If we assume that the *s/a* pattern of the unbound G-stretch is independent of the neighbouring nucleotides, then the probability of having a G-stretch with the *s/a* pattern complementing the triplex is 1/8, since three Gs can adopt $2^3$ = 8 possible *s/a* patterns. The non-native *s/a* patterns may lead to formation of a stable two-quartet GQ with strand slippage. Moreover, even with the correct *s/a* pattern the G-stretch approaching the triplex must land correctly, i.e. every G must approach a G-triad; otherwise locally misfolded structures can form. The G-strand approach is likely affected by the connecting loop. Also locally misfolded G-hairpins can misfold the whole structures, going from misfolded hairpin through misfolded triplex to misfolded GQ. This adds complexity in a form of other parallel pathways. Therefore, any idealized pathway shows only one possibility and in the best case could represent a 'master pathway', along which other misfolded intermediates occur. Taking into account the slowness of the GQ folding, we suggest that only a tiny fraction of the individual molecules reaches the native basin (fold) directly via the master pathway (51,119). Majority of them likely become trapped in different folds rather than staying truly unstructured and need to unfold before taking another folding attempt.

The idealized folding pathway proposed above also does not take into account a non-specific collapse (51), in which denatured oligonucleotides form a 'coil' not resembling the GQ and then slowly by diffusion and random rare transitions progress towards the native or misfolded GQ. In fact, it is even possible that most of the individual oligonucleotide strands undergo some non-specific collapse, rather than following a more-or-less well defined idealized folding pathways with a series of distinct intermediates. Our simulations of four-stranded and triplex intermediates located numerous stable (on the simulation time scale) structures stabilized by non-GQ interactions such as structures with perpendicular hydrogen-bonded G-strands, pairing between loop nucleotides and G-stretches, etc. (60,61). Since these two mechanisms may coexist, we can imagine a scenario in which relatively fast folding pathways exchange intermediates with the slow non-specific processes, allowing diffusion between the different major conformational basins. The nonspecific diffusive processes could help in structuring the propeller loops.

Our last assumption when constructing the pathway was that all the G-hairpins have similar folding rate constants, so their populations are proportionate to their unfolding rates modelled by their relative lifetimes in the simulations. If this is not the case, then even at first sight less stable hairpins might participate in the folding process, provided their folding rate constants are high enough to keep their sufficient concentrations.

**CONCLUDING REMARKS**

Free energy surface of GQ-forming sequences is rugged enough to allow separation of distinct energy minima and studying the corresponding structures independently from the others. G-hairpins were suggested to be the simplest such minima significantly participating in the GQ folding pathways (42,56). We examined their properties by standard MD simulations starting from a diverse set of folded G-hairpins with three $c$WH base pairs and long T-REMD simulation starting from the unfolded state.

We show that antiparallel G-hairpins can easily form in the earliest stages of human telomeric GQ folding to support lateral (both narrow and wide groove) and diagonal loops, with almost barrier-less rearrangements between them (Figure 2). The experimentally estimated microseconds to millisecond folding range (40-42,72) of G-hairpins is consistent with the simulations. The G-hairpins can also participate in many structural transitions in the later stages of the folding.

Despite a clear trend to form the antiparallel structures, there is no predisposition to fold into three-base-pair hairpins that would correspond to direct folding into specific GQ topologies with specific *syn/anti* patterns. Further, rather than forming the optimal three-base-pair G-hairpins, the T-REMD simulation reveals a rich zoo of structures, mixture of hairpins, slipped structures, entirely misfolded structures not resembling hairpins, etc. (Figure 6). Nevertheless, antiparallel structures plausible to move towards various G-triplexes and GQs are sufficiently populated and easily accessible.

Our key result is that an ideal unfolded ensemble of d[GGGTTAGGG] populates all 64 possible *s/a* patterns of the two G-stretches. Thus we predict that an ideal unfolded ensemble of a full human telomeric GQ sequence should populate all 4096 *s/a* combinations of its four G-stretches. Out of the 4096 *s/a* patterns, 2336 are compatible with some three- or a two-quartet GQ; the algorithm is given in the Supporting Information and disregards GQs with G-bulges (120). We suggest that a fully unfolded GQ-forming sequence initially folds into a rich spectrum of structures with no selection of the native *s/a* GQ pattern in the very early stages of folding. There is no sign of any straightforward funnel-like mechanism towards a single GQ topology inherent to the human telomeric sequence. The rich spectrum of various misfolded structures with non-native interactions further supports the view of multi-pathway nature of the GQ folding with the final GQ topology becoming thermodynamically prevalent in a very slow process of countless folding – unfolding events. It is consistent with recent experiments and observed time scales (43), but not consistent with simple single-pathway models.

The complexity of the G-hairpin conformational space gives just a flavour of the intricacy of the conformational space of the full GQ sequence when having unlimited time resolution. This supports the idea of very rough free energy landscape, with only a tiny fraction of structures directly folding towards the native GQ via a fast folding track. Note that even for simple systems such as DNA hairpins with canonical base pairs, a slow folding kinetics compared to ideal semi-flexible polymer was proposed (109,121,122). GQ folding should be incomparably more complex. The GQ folding could be consistent with the kinetic partitioning mechanism (KPM) with a set of deep competing minima on the folding landscape which are well separated from each other, with the native state being one of them (45,119). The number of resolved substates may depend on the resolution of the method. The folding landscape should also include phases of non-specific collapse (NSC) in which the oligonucleotides form a 'coil' not resembling the GQ and then slowly by diffusion and random rare transitions progress towards the native or misfolded GQs. The KPM and NSC mechanisms likely coexist and the real folding process can involve a very complex spectrum of structures with diverse populations and lifetimes, from milliseconds to days. The position of the global minimum can be modulated by details of the sequence and specific experimental conditions.

The simulations demonstrate how the flanking nucleotides affect behaviour of G-hairpins, which may be related to influence of flanking nucleotides on the GQ folds in experiments. Narrow groove hairpins seem to be particularly stable if their first G after the loop is *syn*-oriented because of formation of several stabilizing interactions (Figures 4 and S3). Our data are in good agreement with the experimentally known GQ *s/a* patterns and groove widths they adopt (22), indicating that at least some of the factors affecting the lifetimes of the G-hairpins may be already local native interactions tuning the final relative stabilities of different GQ folds.

In contrast to the antiparallel G-hairpins, parallel G-hairpins with propeller loops are very unstable in the simulations. It indicates that formation of propeller loops during GQ folding is not straightforward, at least compared to the lateral and diagonal loops. We tentatively suggest that the propeller loops could be structured by diffusion and series of consecutive random rare transitions after a nonspecific collapse of the full GQ-forming sequence. For this model, however, we do not have a direct simulation support yet.

The present work has also limitations that are inherent to the MD method. Besides the simulation time, the results are affected by the approximate nature of the force field. Two problems could specifically affect quantitative results of this study, although they are unlikely to influence its basic conclusions. First, the technique may underestimate the stability of the *c*WH GG base-paired G-hairpin stems (61). In this case the population and stability of the fully paired antiparallel G-hairpins would be larger than seen in our simulations, with a smaller role of competing structures. This would further strengthen our basic conclusions. Second, the force field may destabilize the propeller loops. This likely would not change the conclusion that the formation of propeller loops is more difficult relatively to the lateral and diagonal loops, but they could be more accessible compared to the present simulation data.

MD simulations of potential intermediates can be used to suggest idealized folding pathways of GQs, similar to those proposed based on experiments (41,42,116,123). Figure 8 shows such pathway for the basket form of the human telomeric GQ. However, if the GQ folding is a multi-pathway process with only a tiny fraction of the individual molecules folding directly to the native fold, such simplified pathways would not reflect the complexity of the real folding landscapes.

We suggest that combination of smart experimental techniques with sophisticated computational approaches is needed to better understand the GQ folding. Atomistic MD provides interesting insights into selected aspects of the GQ folding. However, complexity of the process will likely require combining MD with other approaches, such as high-resolution coarse-grained modelling (124), systematic conformational searches (125) and perhaps other methods. The folding pathways may vary for different GQs and for different experimental conditions having different unfolded ensembles (45). Thus, while we may be able to understand the basic principles of GQ folding, it is less likely that there exists a uniform mechanism valid for all sequences and folding conditions. The intrinsic GQ folding landscapes can be modulated by other molecules including proteins and, when a GQ-folding sequence needs to be first freed from a double helix, also by the kinetics of the duplex melting and release of the G-stretches. These factors may determine which structures are populated on the biochemically relevant time-scales. Computational approaches could be used to study the effect of the unfolded ensemble on the subsequent folding.

## SUPPLEMENTARY DATA

Supplementary Data are available at NAR online: Supporting Materials and Methods, Supporting Results, Supporting Tables S1 – S7, Supporting Figures S1 – S9, Supporting Information Appendix Tables SA1 – SA24.

## FUNDING

This work was supported by the Czech Science Foundation [grant number P208/11/1822]; the project "CEITEC – Central European Institute of Technology" [CZ.1.05/1.1.00/02.0068] from European Regional Development Fund; the Ministry of Education, Youth and Sports of the Czech Republic [project No. LO1305]; Student Project [IGAPrF_2014023] of Palacký University to P.K, P.B., M.O.; the career development grant from the European Organization for Molecular Biology [IG2535] and the Marie-Curie Re-integration grant (ECOPOD) to L.T.; the European Research Council [Starting Grant S-RNA-S, no. 306662] to G.B.; support by Praemium Academiae to J.S. Funding for open access charge: Czech Science Foundation [grant number P208/11/1822].

**TABLES AND FIGURES**

**Table 1.** Outcome of all 48 simulations (200 ns each) of the antiparallel d[AGGGTTAGGG] hairpin with the bsc0$\chi_{OL4}\varepsilon\zeta_{OL1}$ force field. Final structures are marked in the following way: empty cell or regular font text mean that the final structure resembles the starting structure with at least two native *c*WH GG pairs at the end of the simulation (loss of one base pair is tolerated). '**P:**' stands for any other hairpin-like structure at the end (perturbed structures, see the text). '**U:**' means entire disruption of the hairpin, e.g. unwinding to ss-helix or coil formation. Characteristic structural rearrangements in the simulations are described by the following symbols: 'O' means opening (loss) of a particular GG pair (1 – closing the loop, 2 – middle, 3 – terminal pair), '*t*W' means transition of the GG pair from *c*WH to *t*WW geometry, 'S5' and 'S3' mean strand slippage towards the 5'- or 3'-terminus with base pairing shifted by one base (Figure 3 depicts example of a structure with S5 slippage), 'G' means conversion from initial narrow groove to wide groove or vice versa (see Figure 2). '*' denotes that the particular rearrangement was reversible and the initial *c*WH pairing was restored. Rearrangements shown for the unstable (**U:**) trajectories are always temporary, i.e., we mark those seen before the entire loss of the structure. If more structural changes occurred, they are divided by ';' and ordered chronologically.

| Groove width | *s/a* pattern | | | | Water and ionic parameters | | |
|---|---|---|---|---|---|---|---|
| | 5'-*asa*---*sas*-3' | 5'-*saa*---*ssa*-3' | 5'-*ssa*---*saa*-3' | 5'-*sas*---*asa*-3' | | | |
| Narrow | 3 O | 1,2,3 tW* | 3O*;1,2,3tW* | **P:** 1 O | | NaCl | J&C |
| | 3 O* | 2 tW | | 3 tW* | TIP3P | KCl | J&C |
| | 3 O* | 2 tW | | | | NaCl | Aq. |
| | 3 O* | | | **U:** S5 | | NaCl | J&C |
| | 3 O | | | | SPC/E | KCl | Aq. |
| | | **P:** 1,2,3 tW | 2,3 tW* | **U:** 1,2,3 O | | KCl | Dang |
| Wide | **U:** S5 | **U:** S5 | **U:** 1,2,3 O*; S5 | **U:** 1,2,3 O | | NaCl | J&C |
| | **P:** S5 | 1,2,3 tW | **U:** 1,2,3 O | **P:** 2,3 O* | TIP3P | KCl | J&C |
| | 3 O | **U:** 1,2,3 tW | See text | **U:** 2,3tW*; 1,2,3O | | NaCl | Aq. |
| | **U:** S5 | **P:** 1,2,3 tW | **U:** 1,2,3 O | | | NaCl | J&C |
| | | | **U:** 1,2,3 O | **U:** 1,2,3 O | SPC/E | KCl | Aq. |
| | **U:** 2,3 O; 1tW | | **P:** 1 O; 2,3 tW | **P:** G; 1 O | | KCl | Dang |

[a] J&C: TIP3P-adapted or SPC/E-adapted Joung&Cheatham parameters; Aq: Amber-adapted Aquist Na$^+$ with Dang&Smith Cl$^-$ parameters; Dang: Dang K$^+$ with Dang&Smith Cl$^-$ paremeters

**Table 2.** Outcome of all 48 simulations (200 ns each) of the antiparallel d[AGGGTTAGGGT] hairpin. Length of the simulations is 200 ns. See the legend of the Table 1 for further details.

| Groove width | s/a pattern | | | | Water and ionic parameters | | |
|---|---|---|---|---|---|---|---|
| | 5'-*asa*---*sas*-3' | 5'-*saa*---*ssa*-3' | 5'-*ssa*---*saa*-3' | 5'-*sas*---*asa*-3' | | | |
| Narrow | | **P:** 2,3 tW; 1 O | **P:** 2,3 tW; G* | **P:** 1,2,3 tW | TIP3P | NaCl | J&C |
| | | **P:** 1,2,3 tW | **P:** 1,2,3 tW | **P:** 1,2,3 tW; G* | | KCl | J&C |
| | | **P:** 2,3 tW | **P:** 1,2,3 tW | **P:** G; 1 O | | NaCl | Aq. |
| | | **P:** 2,3 tW | | | SPC/E | NaCl | J&C |
| | | | | | | KCl | Aq. |
| | | **P:** 1,2,3 tW | | **P:** G; 1 O | | KCl | Dang |
| Wide | **U:** 1,2 tW | | **P:** 1,2,3 tW | **U:** S5*; 1,2,3 O | TIP3P | NaCl | J&C |
| | **U:** S5 | **P:** S5 | | **U:** 1,2,3 O | | KCl | J&C |
| | 3 O | **P:** 1,2,3 tW | **U:** S5 | **P:** S5 | | NaCl | Aq. |
| | 3 O | | | **U:** S5 | SPC/E | NaCl | J&C |
| | **U:** S5 | | **P:** S5 | **P:** 1 O | | KCl | Aq. |
| | **U:** 1,2,3 O | | **P:** S5 | **P:** 1,2 O | | KCl | Dang |

[a] J&C: TIP3P-adapted or SPC/E-adapted Joung&Cheatham parameters; Aq: Amber-adapted Aquist Na$^+$ with Dang&Smith Cl$^-$ parameters; Dang: Dang K$^+$ with Dang&Smith Cl$^-$ paremeters

**Table 3.** Outcome of 48 simulations (200 ns each) of the antiparallel d[AGGGTTAGGGA] hairpin. Length of the simulations is 200 ns. See the legend of the Table 1 for further details.

| Groove width | s/a pattern | | | | Water and ionic parameters | | |
|---|---|---|---|---|---|---|---|
| | 5'-*asa*---*sas*-3' | 5'-*saa*---*ssa*-3' | 5'-*ssa*---*saa*-3' | 5'-*sas*---*asa*-3' | | | |
| Narrow | 3 O | **U:** 2,3 O | **U:** 3 O; 1,2,3 tW | **U:** 1,2,3 tW | TIP3P | NaCl | J&C |
| | **P:** 1*,2,3tW;3O* | **U:** 1,2,3 tW; G | 1,2,3 tW* | **U:** 1,2,3 tW | | KCl | J&C |
| | 3 O | | | | | NaCl | Aq. |
| | 3 O | **P:** 2,3 O | | | SPC/E | NaCl | J&C |
| | 3 O | | | | | KCl | Aq. |
| | 3 O | **P:** 1,2,3 tW | 2,3 tW* | **U:** See text | | KCl | Dang |
| Wide | 3 O | 1,2,3 tW* | **P:** S3 | **U:** 1 O | TIP3P | NaCl | J&C |
| | 3 O | **P:** 1,2,3 tW | | S3*; 1*,2*,3O | | KCl | J&C |
| | **U:** 2,3 O | **P:** 1,2,3 tW | **P:** S3 | **P:** S5 | | NaCl | Aq. |
| | 3 O | | | **P:** 1,2 O | SPC/E | NaCl | J&C |
| | 3 O | | **P:** 3 O; S3* | **P:** 1 O | | KCl | Aq. |
| | **U:** 3 O | | **U:** S3 | **U:** 1 O | | KCl | Dang |

[a] J&C: TIP3P-adapted or SPC/E-adapted Joung&Cheatham parameters; Aq: Amber-adapted Aquist Na$^+$ with Dang&Smith Cl$^-$ parameters; Dang: Dang K$^+$ with Dang&Smith Cl$^-$ paremeters

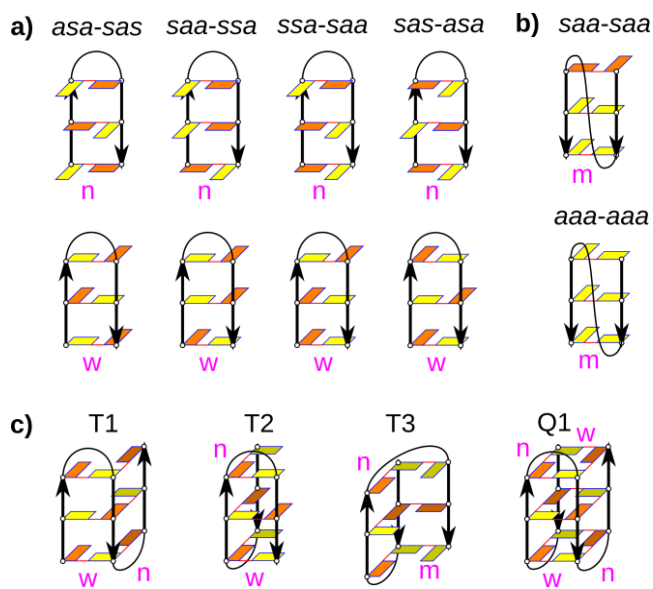

**Figure 1.** Basic schemes of the simulated structures. a) Antiparallel hairpins with four different *syn/anti* (*s/a*) patterns. Each pattern has one variant with narrow groove and one with wide groove. b) Parallel stranded duplexes. c) Triplexes and a chair-like quadruplex with different loop types and order, all with *s/a* patterns capable of folding to the basket-type quadruplex 143D without any *syn ↔ anti* transition. Deoxyguanosine residues are shown as rectangles, *syn*-oriented are in orange, *anti*-oriented in yellow. Black lines depict backbone and arrows mark the 5'-end-to-3'-end direction. Red lines mean hydrogen bonding. 'n', 'w' and 'm' stand for narrow, wide and medium groove, respectively. Loop and flanking residues are not shown. The base pairing of GG pairs is *c*WH. For the sake of brevity, *s/a* pattern abbreviations in the Figure are compacted, e.g. *asa-sas* instead of 5'-*asa*---*sas*-3' used in the text.

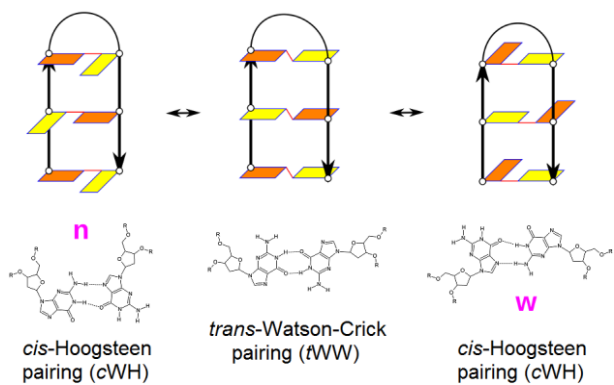

**Figure 2.** Three types of antiparallel hairpins: narrow groove lateral loop hairpin with *c*WH pairs (left), diagonal hairpin with *t*WW pairs (middle) and wide groove lateral loop hairpin with *c*WH pairs (right). The structural schemes are visualized as in the Figure 1.

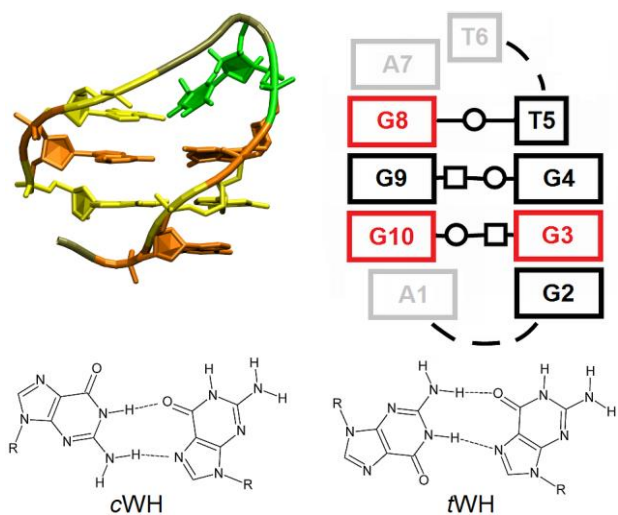

**Figure 3**. Antiparallel d[AGGGTTAGGG] hairpin slipped towards its 5'-terminus by one residue (this direction of slippage is marked as S5 in the text). (top left) Atomistic structure: *syn*-oriented Gs are orange, *anti*-oriented Gs yellow, the first thymine in the loop is green and the backbone of other loop residues is tan. (top right) Secondary structure scheme. Red and black colours correspond to *anti* and *syn* conformation, respectively. Gray residues are not depicted in the atomistic structure. The two GG base pairs are non-native *t*WH, the GT base pair is *t*WW. (bottom) Comparison of *c*WH and *t*WH base pairs.

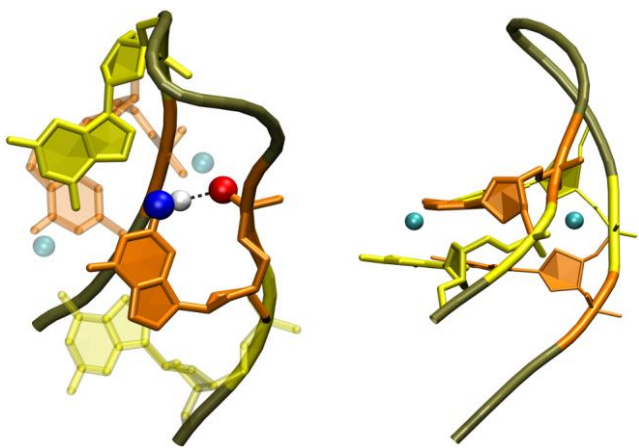

**Figure 4**. (left) Intra-nucleotide base-phosphate hydrogen bonding between amino and phosphate groups of *syn*-oriented G following the lateral TTA loop in the narrow groove hairpins. (right) Two dominant cation binding sites of narrow groove hairpins associated with the *syn-anti* GpG dinucleotide step: near O6 carbonyl oxygens and in the narrow groove. The ions are also visible in the left part as transparent spheres. *Syn*-oriented Gs are in orange, *anti*-oriented Gs in yellow, backbone of other residues is tan and the cations are cyan.

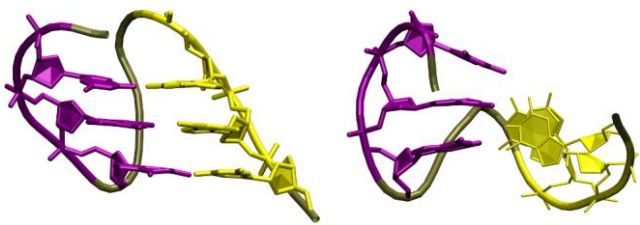

**Figure 5.** (left) Native parallel hairpin with *c*WH GG base pairs. (right) Typical intermediate in the counter-rotation unfolding mechanism. The two G-stretches (purple and yellow) are nearly perpendicular and the structure of the V-shaped propeller loop is already lost. Backbone of other residues is tan.

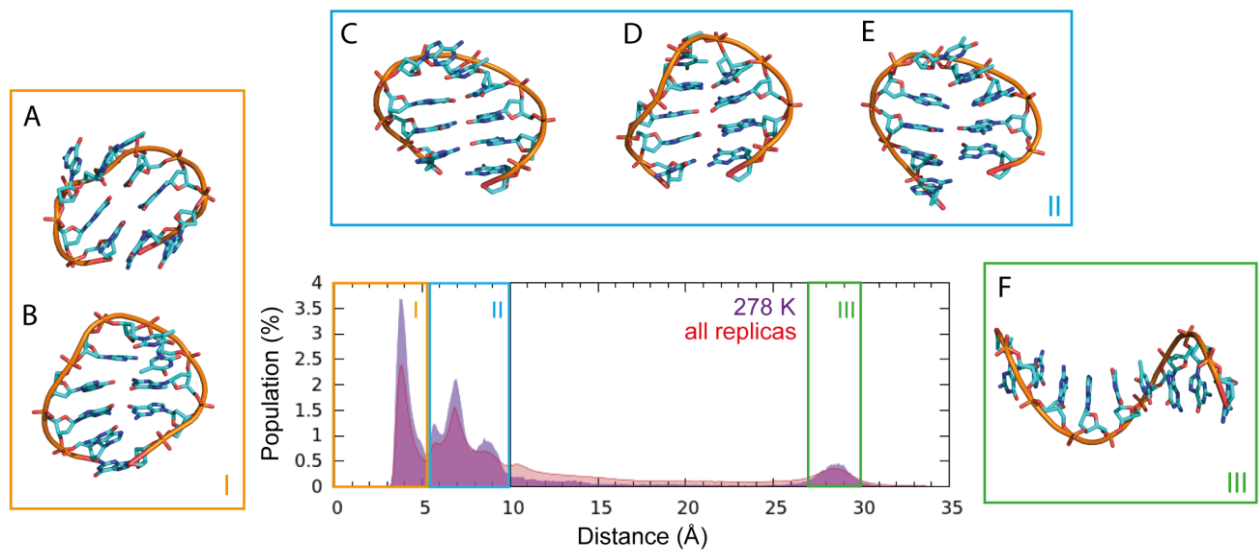

**Figure 6**. T-REMD simulation of d[GGGTTAGGG]. The population of end-to-end distance represented by a distance between centres of masses of terminal guanine bases in the reference 278 K replica and in all 64 replicas. The region I with end-to-end distance bellow 5 Å corresponds to structures with stacked terminal guanines, region II includes native conformation with terminal GG base pair as well as the other bent conformations with strand slippage, and region III corresponds to the straight ssDNA. **A**: circular structure containing G3-G9 and T4-G8 base pairs and stacked terminal guanines; **B**: hairpin containing G2-G8 and G3-G7 base pairs and stacked terminal guanines; **C**: hairpin with G2-G9, G3-G8, and T4-G7 base pairs; **D**: hairpin with G1-G9, G2-G8, and G3-G7 base pairs; **E**: hairpin with G1-G8, G2-G7, and G3-A6 base pairs; **F**: unfolded ss-helix structure.

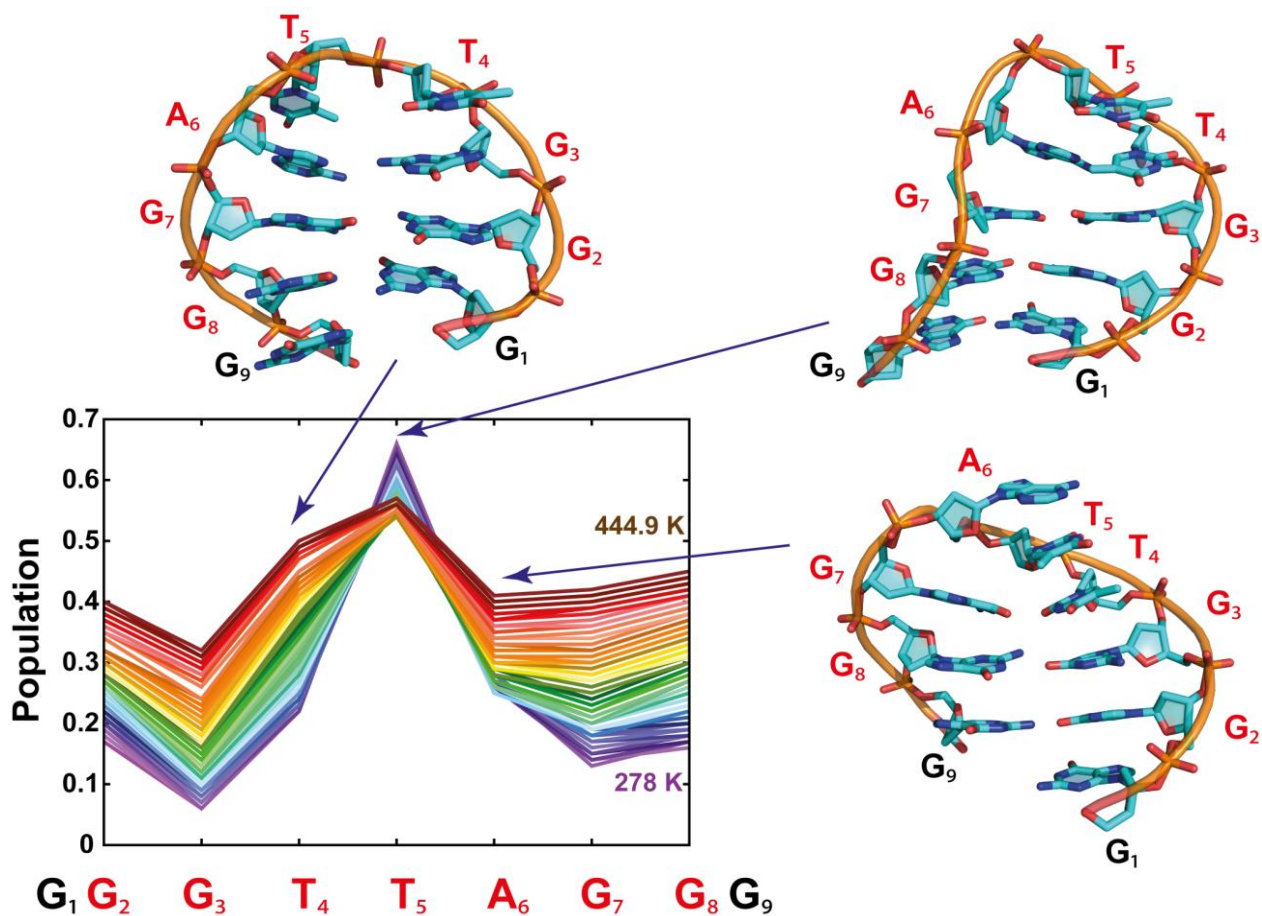

**Figure 7.** The bending propensity at different nucleotides along the d[GGGTTAGGG] sequence observed in the T-REMD simulation. The bent state was defined by N-1(C1')…N(C1')…N+1(C1') angle below 120° (see Supporting Information for further details). The lines correspond to replicas following temperature and are coloured according to the growing temperature from violet to maroon. The y-axis represents the population of the bend located at a particular nucleotide (x-axis) averaged over the entire replica. The representative structures taken from the T-REMD simulation correspond to the most populated bent conformations.

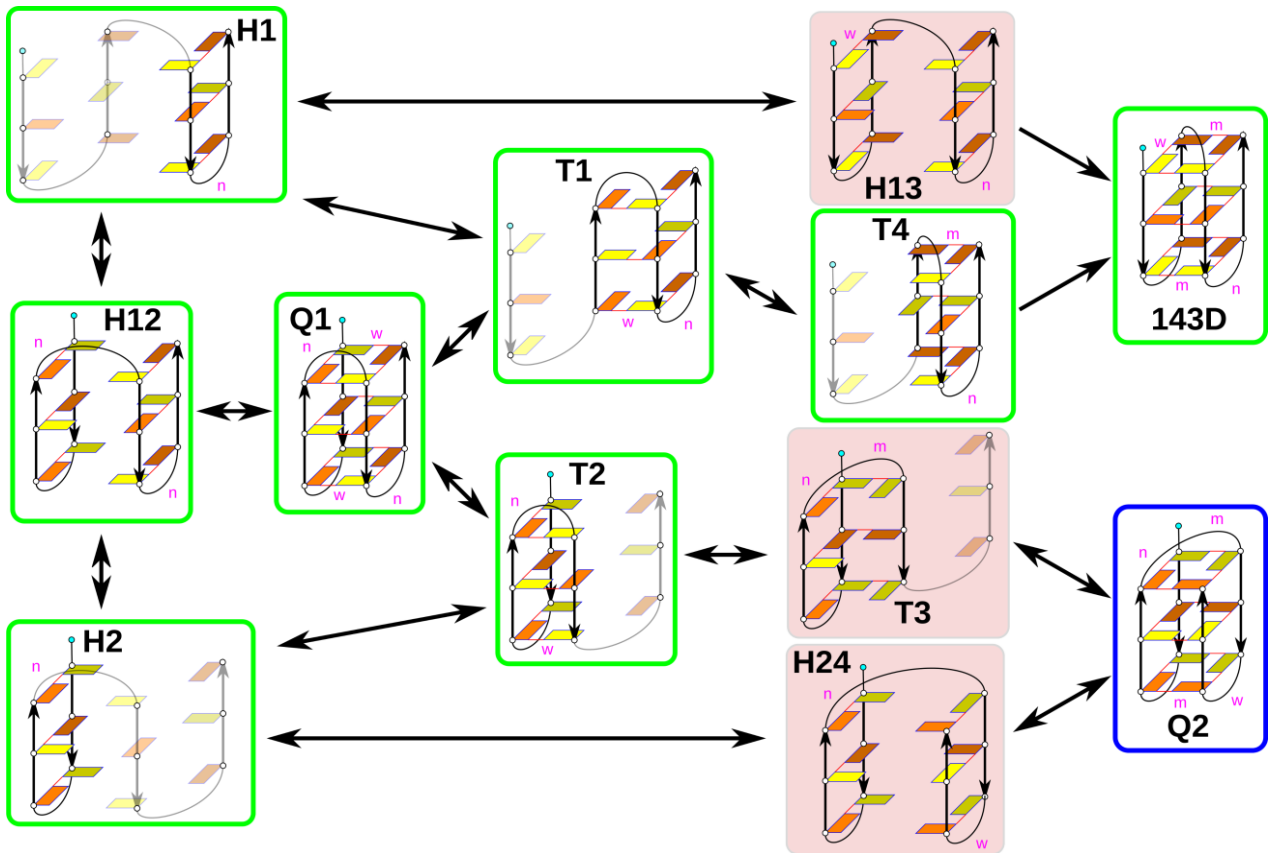

**Figure 8**. Plausible idealized folding pathway of the basket-type GQ 143D. The structural schemes are visualized as in the Figure 1 and the flanking 5'-terminal adenine is depicted as a cyan circle. The green squares indicate structures that we found stable in simulations and the native basket-type GQ 143D. The blue square depicts the alternative basket fold **Q2** with the same *syn/anti* pattern. The red squares indicate intermediates which we suggest are less likely to participate in the folding pathway. Those parts of the **H1**, **H2** and **T1-T4** structures that were not directly simulated are transparent in the schemes. Stability of the **H12**, **H13** and **H24** structures is assessed based on simulation behaviour of their separate G-hairpin parts. The model suggests a lack of a suitable pathway to the alternative topology **Q2**.